\documentclass[a4paper,11pt]{article}
\usepackage{pos}
\usepackage{amsmath}
\usepackage{relsize}
\usepackage{graphicx}
\usepackage{float}
\usepackage{comment}
\usepackage[utf8]{inputenc}
\title{Recent Highlights from the STAR Experiment}

\author*[a]{Rutik Manikandhan}

\onbehalf{for the STAR collaboration}

\affiliation[a]{University of Houston,\\
  4302 University Dr, Texas, U.S.A,}

\emailAdd{manikandhan.rutik@gmail.com}

\abstract{We summarize the latest correlation and fluctuation measurements derived from the RHIC Beam Energy Scan-II (BES-II) data, collected by the STAR experiment. We will focus on the recent results of higher-order net-proton cumulants ($C_{1}-C_{4}$) in the energy range of 7.7 to 27 GeV Au+Au collisions.
Furthermore, we will present measurements of transverse momentum correlations of charged particles, particularly focusing on 2-particle correlators and their dependency on centrality in 3.0 and 3.2 GeV
Au+Au collisions.}

\FullConference{%
  FAIR Next Generation of Scientists,\\
  23-27 September 2024\\
  Hotel Medena, Donji Seget, Croatia
}

\begin{document}
\maketitle

\section{Introduction}
The study of event-by-event correlations and fluctuations in global quantities can provide insight into the properties of the hot and dense matter created in
Au+Au collisions at ultra relativistic collision energies \cite{PhysRevC.66.044904} \cite{PhysRevC.63.064904}.
The event-by-event fluctuations of conserved quantities such as net charge, net-baryon number, and
net strangeness are predicted to depend on the non-equilibrium correlation length, $\xi$, and thus serve as indicators of critical behavior \cite{Stephanov_1998} \cite{PhysRevD.60.114028}. \newline
Correlations of transverse momentum, $p_{t}$, have been proposed as a measure of thermalization \cite{PhysRevLett.92.162301} \cite{PhysRevC.85.014905} and as a probe for the critical point of quantum chromodynamics (QCD) \cite{PhysRevD.65.096008} \cite{PhysRevLett.102.032301}. Studying these observables and quantifying their deviations from baselines of uncorrelated emissions could elucidate the possible existence of the critical point.

In these proceedings, we present the results of proton cumulant measurements $(C_{1}-C_{4})$ and factorial measurements $(\kappa_{1}-\kappa_{4})$ along with their ratios from Beam Energy Scan-II (BES-II) collider energies at $\sqrt{s_{NN}}$ = 7.7, 9.2, 11.5, 14.6, 17.3, 19.6 and 27 GeV and preliminary results of the 2-particle transverse momentum correlators from the BES-II Fixed-Target (FXT) energies at $\sqrt{s_{NN}}$ = 3.0 and 3.2 GeV and how they probe the QCD phase digram for the existence of a critical point.

\section{Analysis Method}
Experimentally measured proton multiplicity distributions are described by the central moments, i.e., $ \langle (\delta N)^{2}\rangle$, $ \langle (\delta N)^{3}\rangle$, $ \langle (\delta N)^{3}\rangle$ etc... The symbol $\langle...\rangle$ indicates the average
for all events, N is the multiplicity of protons in a given
event, and $\delta N = N - \langle N \rangle$ is the deviation. The relations between the cumulants $C_{n}$ and the central moments are defined as:

\[
\begin{aligned}
\text{Mean:} \quad & M = \langle N \rangle = C_1 \\
\text{Variance:} \quad & \sigma^2 = \langle (\delta N)^2 \rangle = C_2 \\
\text{Skewness:} \quad & S =  \langle (\delta N)^{3} \rangle /\sigma^{3} = C_{3}/C_{2}^{3/2}\\
\text{Kurtosis:} \quad & \kappa =  \langle (\delta N)^{4} \rangle/\sigma^{4} - 3 = C_{4}/C_{2}^{2}\\
\end{aligned}
\]

The ratios of the cumulants are often used to reduce volume dependence: $C_{2}/C_{1} = \sigma^{2}/M$ , $C_{3}/C_{2} = S\sigma$, and $C_{4}/C_{2} = \kappa \sigma^{2}$. An additional advantage is that the ratios of these cumulants can be readily compared with theoretical calculations of susceptibility  ratios for, e.g. $\sigma^{2}/M = \chi_{2}/\chi_{1}$, $S\sigma = \chi_{3}/\chi_{2},$ and $\kappa \sigma^{2} = \chi_{4}/\chi_{2}$. In case there are no intrinsic correlations among the measured particles, all ratios of the cumulants are unity, thus Poisson statistics is a trivial baseline for experimentally measured cumulant ratios \cite{Ejiri_2006} \cite{PhysRevLett.109.192302}. \newline

All of the shown measurements have been corrected for pile-up, detector efficiency and centrality bin width effects which have been discussed in detail in \cite{thestarcollaboration2025precisionmeasurementnetprotonnumber}, \cite{PhysRevC.107.024908}, \cite{PhysRevC.104.024902} and references therein. \newline

$(p_{t})$ correlations are characterized by the two-particle correlation function
defined as the covariance:

\begin{equation}
    \langle \Delta p_{t,i}, \Delta p_{t,j} \rangle = \frac{1}{N_{events}} \mathlarger{\mathlarger{\sum}}_{k=1}^{N_{events}} \frac{C_{k}}{N_{k}(N_{k}-1)}
\end{equation}

where

\begin{equation}
    C_{k} = \mathlarger{\mathlarger{\sum}}_{i=1}^{N_{k}}\mathlarger{\mathlarger{\sum}}_{j=1,j\neq i}^{N_{k}} (p_{t,i} - \langle \langle p_{t} \rangle \rangle)(p_{t,j} - \langle \langle p_{t} \rangle \rangle)
\end{equation}

$N_{events}$ is the number of events, $N_{k}$ is the number of
tracks in the $k^{th}$ event, and $p_{t,i}$ is the transverse momentum of the $i^{th}$ track in the given event. The event averaged $p_{t}$ is defined as:

\begin{equation}
    \langle \langle p_{t} \rangle \rangle  = \frac{\sum_{k=1}^{N_{events}}\langle p_{t} \rangle_{k}}{N_{events}}
\end{equation}

where $\langle p_{t} \rangle_{k}$ is the average $p_{t}$ of the $k^{th}$ event defined as:

\begin{equation}
    \langle p_{t} \rangle_{k} = \frac{\sum_{i=1}^{N_{k}}p_{t,i}}{N_{k}}
\end{equation}

To characterize, two-particle $p_{t}$ correlations, we present the relative dynamical correlation, $\sqrt{\langle \Delta p_{t,i} \Delta p_{t,j} \rangle}/\langle \langle p_{t} \rangle \rangle$. It represents the magnitude of the dynamic fluctuations of the average transverse momentum in units of $\langle \langle p_{t} \rangle \rangle$. This scaling cancels out detector efficiency effects \cite{PhysRevC.99.044918} and flow effects \cite{PhysRevLett.92.162301} making it an ideal probe for critical point searches.

\section{Results and Discussion}

\begin{figure}[ht]
\centering
\includegraphics[width=\linewidth]{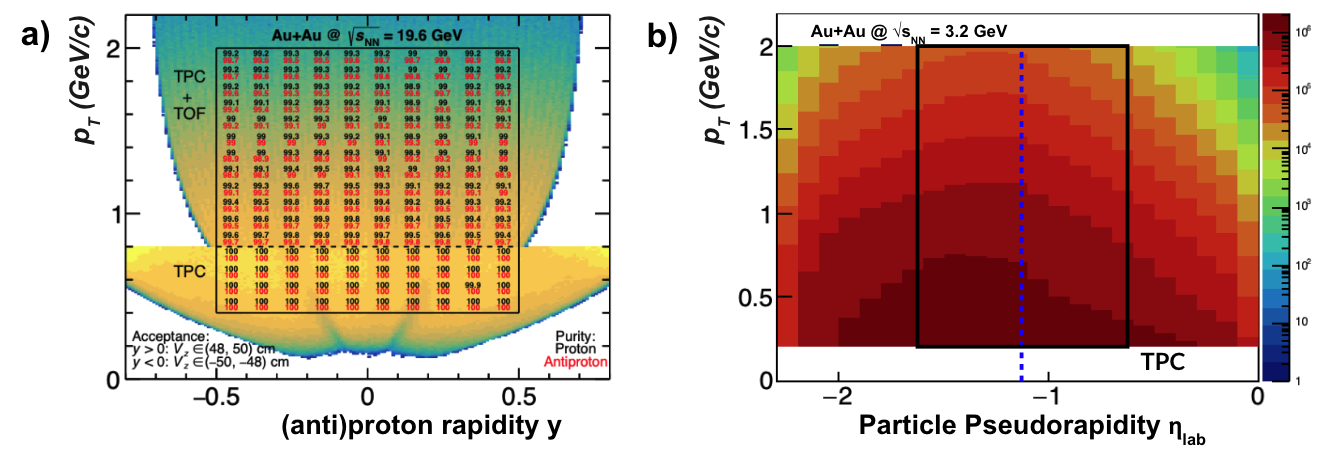}
\caption{\label{fig:Proton_acceptance} (Anti)Proton acceptance along with purity at $\sqrt{s_{NN}} =$ 19.6 GeV and charged particle acceptance at $\sqrt{s_{NN}} =$ 3.2 GeV, the dashed blue line is mid-pseudorapidity at $\sqrt{s_{NN}} =$ 3.2 GeV.}
\end{figure}

The results shown here for the proton cumulants are within a common kinematic acceptance across all energies. The Time-Projection Chamber (TPC) and Time-Of-Flight detectors have been used for identifying the protons and the anti-protons. The TPC identifies the low $p_{T}$ ($0.4 < p_{T} < 0.8$ GeV/c ) protons and anti-protons with high purity and the TOF identifies particles at higher $p_{T}$ ($0.8 < p_{T} < 2.0$ GeV/c ) \cite{PhysRevC.107.024908} within a rapidity window ($|y| < 0.5 $) as shown in Fig. \ref{fig:Proton_acceptance} a).

The STAR detector had major upgrades done for BES-II, which allowed measurements of charged particles at wider pseudorapidity acceptances ($|\eta| < 1.6$). This allowed for a new centrality definition, namely RefMult3X and due to larger multiplicity within the acceptance we have better centrality resolution.  

The STAR detector has recorded data in the Fixed-Target mode as well, this allowed data taking at even lower energies, all the way down to $\sqrt{s_{NN}} $ =  3.0 GeV. For the analysis of $(p_{T})$ correlations, all charged particles in an acceptance of $p_{T}$ ($0.20 < p_{T} <2.0 $ GeV/c)  and $\eta_{cm}$ ($|\eta_{cm}| < 0.5$), where $\eta_{cm} = \eta_{lab} - \eta_{mid}$ (black box in Fig.\ref{fig:Proton_acceptance} b)) are analyzed and compared as a function of collision energy.

\begin{figure}[ht!]
\centering
\includegraphics[width=\linewidth]{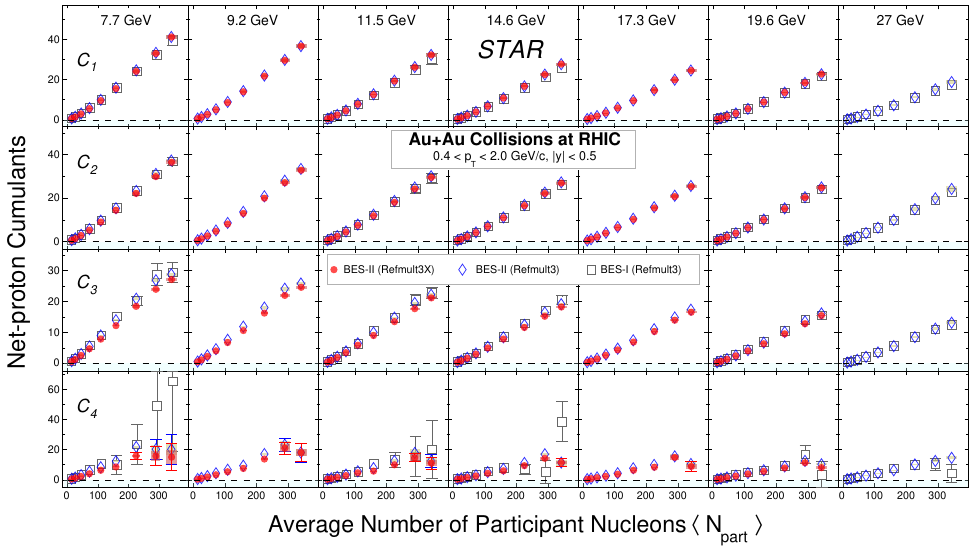}
\caption{\label{fig:Proton_cumulantratios}  
Cumulants of net-proton multiplicity distribution from $\sqrt{s_{NN}}$ = 7.7$-$27 GeV as a function of collision centrality ($\langle N_{\text{part}} \rangle$) in Au+Au collisions at STAR$-$RHIC. Results from BES$-$II with RefMult3X (RefMult3) used for centrality definition are shown as red (blue) markers, while those from BES$-$I \cite{PhysRevC.104.024902} (RefMult3) are shown as open squares. The bars and bands on the data points from BES$-$II represent statistical and systematic uncertainties, respectively. Total uncertainties are shown for BES$-$I data points as bars on data points.}

\end{figure}

Fig.\ref{fig:Proton_cumulantratios} shows how the net-proton cumulants ($C_{2}/C_{1}, C_{3}/C_{2}, C_{4}/C_{2}$) depend on centrality with two different centrality definitions for BES-II and these measurements are compared to previous BES-I measurements \cite{PhysRevC.104.024902}.

The results are consistent with  previous BES-I measurements. The cumulant ratios have a smooth variation across centrality and collision energy, and higher centrality resolution is observed to improve the ratios.

\begin{figure}[ht!]
\centering
\includegraphics[width=0.65\linewidth]{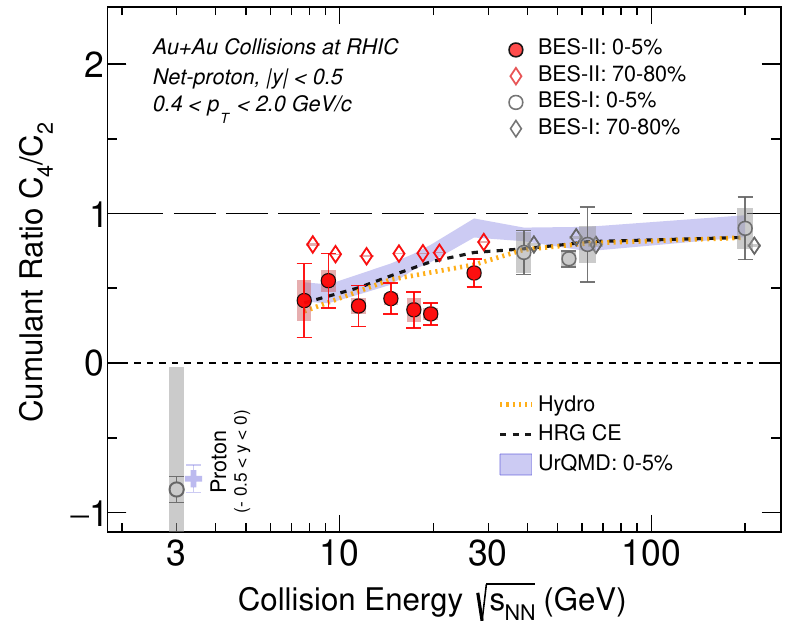}
\caption{\label{fig:Proton_kurtosis} Collision energy dependence of net-proton $C_{4}/C_{2}$ in 0-5$\%$ and 70-80$\%$ centrality classes. Theoretical calculations from a hydrodynamical model \cite{Braun-Munzinger:2020jbk} (Hydro, orange dashed line), thermal model with canonical treatment for baryon charge \cite{PhysRevC.105.014904} (HRG CE, black dashed line), transport model \cite{Bleicher_1999} (UrQMD, violet band) are also presented.}
\end{figure}

The collision energy dependence of net-proton ($C_{4}/C_{2}$) in $0-5\%$ centrality class is shown in Fig.\ref{fig:Proton_kurtosis}. Compared to various non-CP model calculations \cite{Braun-Munzinger:2020jbk},\cite{PhysRevC.105.014904},\cite{Bleicher_1999} and data in 70-80\% peripheral collisions,  the net-proton $C_{4}/C_{2}$ measurement in 0-5\% collisions shows a minimum around $\sqrt{s_{NN}} $ = 19.6 GeV for significance of deviation at $\sim$ 2–5 $\sigma$.

\begin{figure}[H]
\centering
\includegraphics[width=0.73\linewidth]{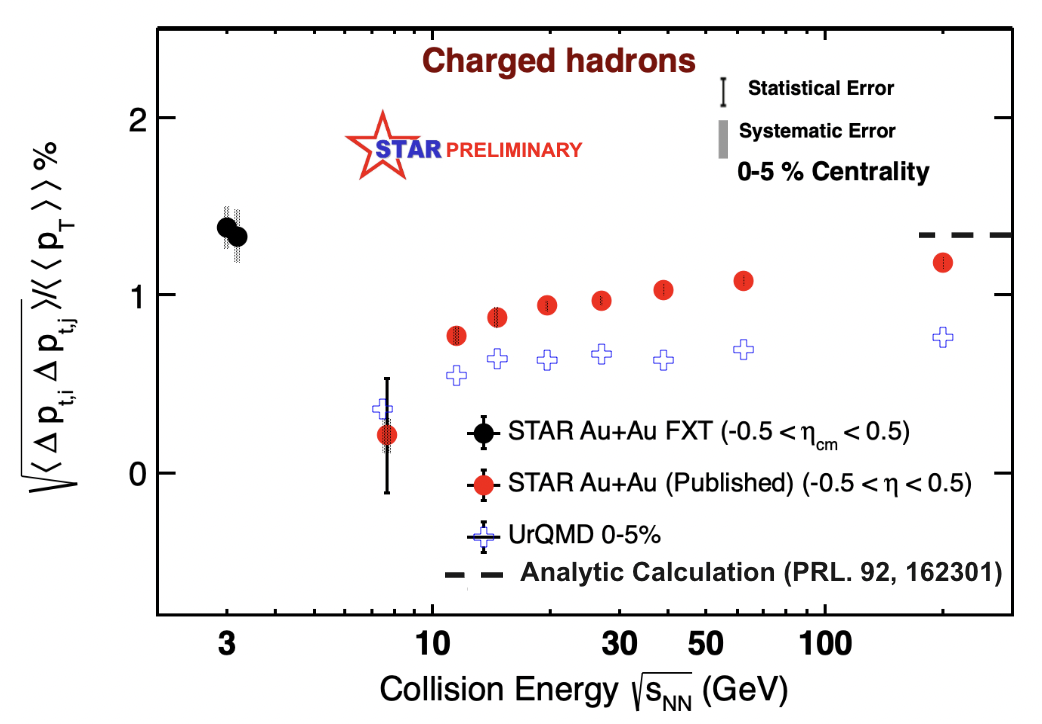}
\caption{\label{fig:pT_Correlator} Collision energy dependence of $\sqrt{\langle \Delta p_{t,i} \Delta p_{t,j} \rangle}/\langle \langle p_{t} \rangle \rangle$ in 0-5$\%$ centrality class. Theoretical calculations from transport model \cite{Bleicher_1999} (UrQMD,blue markers) and analytic method \cite{PhysRevLett.92.162301} (dashed line) are shown }
\end{figure}
\vspace{-1em}

The collision energy dependence of the 2-particle $(p_{T})$ correlator in most central collisions is shown in Fig.\ref{fig:pT_Correlator}, a breaking of monotonicity is observed in the high-baryon density region. This non-monotonous dependence on collision energy could possibly be described due to critical phenomena and the correlation length changing as a function of collision energy \cite{PhysRevLett.92.162301}. 

The dashed line shown in Fig. \ref{fig:pT_Correlator} is a theoretical calculation \cite{PhysRevLett.92.162301} for the expected baseline if the system is thermalized and correlation length remains constant.

We see that the transport code calculations \cite{Bleicher_1999} always deviate from our measurements at the collider energies but qualitatively capture the trend.

\section{Conclusion}
In these proceedings, BES-II measurements of net proton cumulants at  $\sqrt{s_{NN}}$ = 7.7, 9.2, 11.5, 14.6, 17.3, 19.6 and 27 GeV and preliminary results of the 2-particle transverse momentum correlators from the BES-II Fixed-Target (FXT) energies at $\sqrt{s_{NN}}$ = 3.0 and 3.2 GeV are presented.\newline
The net-proton cumulants are discussed as a function of centrality and collision energies, they are compared to the previous measurements from BES-I and to various non-CP models to quantify any kind of critical behavior.
The net-proton $C_{4}/C_{2}$ measurement in 0-5\% collisions shows a minimum around $\sqrt{s_{NN}} $ = 19.6 GeV for significance of deviation at $\sim$ 2–5 $\sigma$. The ($p_{T}$) correlations exhibit a non-monotonic dependence on collision energy in central collisions, potentially signaling critical phenomena. Ongoing studies aim to explore additional energy ranges in this regime and establish theoretical baselines to better quantify any observed deviations.\newline
These highlights underscore the importance of additional measurements from STAR’s fixed-target (FXT) energy program to confirm the existence and pinpoint the location of the critical point on the QCD phase diagram.

\section{Acknowledgements}
This work was supported by various international funding agencies including those from the U.S., China, Europe, Korea, Japan, and Chile. We also acknowledge the computing resources from BNL, LBNL, and the Open Science Grid.
\bibliographystyle{JHEP}
\bibliography{biblio}

\providecommand{\href}[2]{#2}\begingroup\raggedright\begin{thebibliography}{10}

\bibitem{PhysRevC.66.044904}
C.~Pruneau, S.~Gavin and S.~Voloshin, \emph{Methods for the study of particle production fluctuations}, \href{https://doi.org/10.1103/PhysRevC.66.044904}{\emph{Phys. Rev. C} {\bfseries 66} (2002) 044904}.

\bibitem{PhysRevC.63.064904}
H.~Heiselberg and A.D.~Jackson, \emph{Anomalous multiplicity fluctuations from phase transitions in heavy-ion collisions}, \href{https://doi.org/10.1103/PhysRevC.63.064904}{\emph{Phys. Rev. C} {\bfseries 63} (2001) 064904}.

\bibitem{Stephanov_1998}
M.~Stephanov, K.~Rajagopal and E.~Shuryak, \emph{Signatures of the tricritical point in qcd}, \href{https://doi.org/10.1103/physrevlett.81.4816}{\emph{Physical Review Letters} {\bfseries 81} (1998) 4816–4819}.

\bibitem{PhysRevD.60.114028}
M.~Stephanov, K.~Rajagopal and E.~Shuryak, \emph{Event-by-event fluctuations in heavy ion collisions and the qcd critical point}, \href{https://doi.org/10.1103/PhysRevD.60.114028}{\emph{Phys. Rev. D} {\bfseries 60} (1999) 114028}.

\bibitem{PhysRevLett.92.162301}
S.~Gavin, \emph{Traces of thermalization from ${p}_{t}$ fluctuations in nuclear collisions}, \href{https://doi.org/10.1103/PhysRevLett.92.162301}{\emph{Phys. Rev. Lett.} {\bfseries 92} (2004) 162301}.

\bibitem{PhysRevC.85.014905}
S.~Gavin and G.~Moschelli, \emph{Fluctuation probes of early-time correlations in nuclear collisions}, \href{https://doi.org/10.1103/PhysRevC.85.014905}{\emph{Phys. Rev. C} {\bfseries 85} (2012) 014905}.

\bibitem{PhysRevD.65.096008}
M.~Stephanov, \emph{Thermal fluctuations in the interacting pion gas}, \href{https://doi.org/10.1103/PhysRevD.65.096008}{\emph{Phys. Rev. D} {\bfseries 65} (2002) 096008}.

\bibitem{PhysRevLett.102.032301}
M.A.~Stephanov, \emph{Non-gaussian fluctuations near the qcd critical point}, \href{https://doi.org/10.1103/PhysRevLett.102.032301}{\emph{Phys. Rev. Lett.} {\bfseries 102} (2009) 032301}.

\bibitem{Ejiri_2006}
S.~Ejiri, F.~Karsch and K.~Redlich, \emph{Hadronic fluctuations at the qcd phase transition}, \href{https://doi.org/10.1016/j.physletb.2005.11.083}{\emph{Physics Letters B} {\bfseries 633} (2006) 275–282}.

\bibitem{PhysRevLett.109.192302}
A.~Bazavov, H.-T.~Ding, P.~Hegde, O.~Kaczmarek, F.~Karsch, E.~Laermann et~al., \emph{Freeze-out conditions in heavy ion collisions from qcd thermodynamics}, \href{https://doi.org/10.1103/PhysRevLett.109.192302}{\emph{Phys. Rev. Lett.} {\bfseries 109} (2012) 192302}.

\bibitem{thestarcollaboration2025precisionmeasurementnetprotonnumber}
T.S.~Collaboration, \emph{Precision measurement of (net-)proton number fluctuations in au+au collisions at rhic},  2025.

\bibitem{PhysRevC.107.024908}
{\scshape The STAR Collaboration} collaboration, \emph{Higher-order cumulants and correlation functions of proton multiplicity distributions in $\sqrt{{s}_{NN}}=3 \mathrm{GeV} \mathrm{Au}\text{+}\mathrm{Au}$ collisions at the rhic star experiment}, \href{https://doi.org/10.1103/PhysRevC.107.024908}{\emph{Phys. Rev. C} {\bfseries 107} (2023) 024908}.

\bibitem{PhysRevC.104.024902}
{\scshape STAR Collaboration} collaboration, \emph{Cumulants and correlation functions of net-proton, proton, and antiproton multiplicity distributions in $\mathrm{Au}+\mathrm{Au}$ collisions at energies available at the bnl relativistic heavy ion collider}, \href{https://doi.org/10.1103/PhysRevC.104.024902}{\emph{Phys. Rev. C} {\bfseries 104} (2021) 024902}.

\bibitem{PhysRevC.99.044918}
{\scshape STAR Collaboration} collaboration, \emph{Collision-energy dependence of ${p}_{t}$ correlations in au + au collisions at energies available at the bnl relativistic heavy ion collider}, \href{https://doi.org/10.1103/PhysRevC.99.044918}{\emph{Phys. Rev. C} {\bfseries 99} (2019) 044918}.

\bibitem{Braun-Munzinger:2020jbk}
P.~Braun-Munzinger, B.~Friman, K.~Redlich, A.~Rustamov and J.~Stachel, \emph{{Relativistic nuclear collisions: Establishing a non-critical baseline for fluctuation measurements}}, \href{https://doi.org/10.1016/j.nuclphysa.2021.122141}{\emph{Nucl. Phys. A} {\bfseries 1008} (2021) 122141} [\href{https://arxiv.org/abs/2007.02463}{{\ttfamily 2007.02463}}].

\bibitem{PhysRevC.105.014904}
V.~Vovchenko, V.~Koch and C.~Shen, \emph{Proton number cumulants and correlation functions in au-au collisions at $\sqrt{{s}_{NN}}=7.7$--200 gev from hydrodynamics}, \href{https://doi.org/10.1103/PhysRevC.105.014904}{\emph{Phys. Rev. C} {\bfseries 105} (2022) 014904}.

\bibitem{Bleicher_1999}
M.~Bleicher, E.~Zabrodin, C.~Spieles, S.A.~Bass, C.~Ernst, S.~Soff et~al., \emph{Relativistic hadron-hadron collisions in the ultra-relativistic quantum molecular dynamics model}, \href{https://doi.org/10.1088/0954-3899/25/9/308}{\emph{Journal of Physics G: Nuclear and Particle Physics} {\bfseries 25} (1999) 1859–1896}.

\end{thebibliography}\endgroup

\end{document}